\documentclass{article}
\usepackage{spconf,amsmath,graphicx,subfigure,float, color}

\ninept

\title{A HYBRID NEURAL NETWORK FRAMEWORK AND APPLICATION TO RADAR AUTOMATIC TARGET RECOGNITION}
\vspace{-2em}
\name{Zhe Zhang, Xiang Chen, Zhi Tian}
\vspace{-2em}
\address{George Mason University, Electrical and Computer Engineering Department  \\
	4400 University Dr, Fairfax, VA 22030, USA. 
	\{zzhang18, xchen26, ztian1\}@gmu.edu 
	\thanks{This work was partly supported by the NSF grants \#1527396 and \#1741338.}}
\begin{document}
%\ninept
%
\maketitle
%{\tiny }
\vspace{-3em}
\begin{abstract}
Deep neural networks (DNNs) have found applications in diverse signal processing (SP) problems. Most efforts either directly adopt the DNN as a black-box approach to perform certain SP tasks without taking into account of any known properties of the signal models, or insert a pre-defined SP operator into a DNN as an add-on data processing stage. This paper presents a novel hybrid-NN framework in which one or more SP layers are inserted into the DNN architecture in a coherent manner to enhance the network capability and efficiency in feature extraction. These SP layers are properly designed to make good use of the available models and properties of the data. The network training algorithm of hybrid-NN is designed to actively involve the SP layers in the learning goal, by simultaneously optimizing both the weights of the DNN and the unknown tuning parameters of the SP operators. The proposed hybrid-NN is tested on a radar automatic target recognition (ATR) problem. It achieves high validation accuracy of 96\% with 5,000 training images in radar ATR. Compared with ordinary DNN, hybrid-NN can markedly reduce the required amount of training data and improve the learning performance.  
%The proposed method also retains the benefits of DNN including learning unknowns from data and doing typical learning jobs such as automatic classification. 
\end{abstract}
\vspace{-.5em}
\section{Introduction}
\label{sec:intro}
\vspace{-.5em}
During the recent decade, deep learning technology, particularly deep neural network (DNN), has gained tremendous popularity in various fields including signal processing (SP) \cite{lecun2015deep, deng2014deep, cochocki1993neural, yu2011deep}. %Meanwhile, signal processing (SP) society is continuously widening its scope and DL/DNN based techniques have been introduced to SP problems such as speech, audio, image and video processing \cite{}. 
%These approaches utilized the benefit of DNN in terms of learning unknown
%but overlooked the information of data model which can also be utilized.
As a data-driven framework, %DNN works for problems with complex and large scale models. Basically, 
DNN treats the learning problem as a ``black-box'' that extracts useful features directly from data. With sufficient training, DNN does not rely on any special structure or property of the processed data, making it universally applicable to diverse problem models. As such, DNN can help to expand the functionality of SP to handle problems that cannot be well-modeled, such as automatic target recognition (ATR) in radar \cite{morgan2015deep, chen2016target, wilmanski2016complex}.

However, improved universality may lead to worsened specialty, that is, a universally good solution is often non-optimal in terms of either accuracy or efficiency. Specializing to the SP field, there are abundant highly-structured or man-made signals with known properties, such as low-rankness or sparsity. DNN, as a data-driven approach which is blind to specialized signal structures, is obviously not as efficient in processing and extracting useful information from those signals with known models and properties. In contrast, traditional SP methods are typically crafted to gainfully utilize such prior knowledge. It is desired to combine SP and DNN judiciously so as to benefit from both sides. 

Most efforts on combining SP with DNN fall under two categories. One is to treat a DNN as a module and insert it into the conventional SP framework to handle some subtasks \cite{metzler2018prdeep}. Conversely, the other is to attach some SP modules to the DNN framework, either before/after the network as a pre-/post-processing stage, or inside the network to perform some specific data processing  \cite{PCAnet, Optnet}. 
Both approaches can be successful in demonstrating the power of DNN in learning features from complex models and validating the efficiency of SP in dealing with structured data. Common to these approaches, the adopted SP operators are pre-defined as ``hyper-parameters'' of the learning problem, in the sense that all design parameters of the SP operators have to be known {\em a priori} and fixed during training. Unfortunately, adopting pre-defined SP operators encounters two major challenges. First, it can be difficult to pre-define some SP operators because of the difficulty in setting its design parameters since they may be unknown or cannot be estimated accurately. 
Second, in the real world, the properties and structural models of the processed data can be partially known only, in the form of a ``gray-box''. Hence, we wish to  not only embed some appropriate SP modules into the DNN to effectively utilize those partially available data properties, but also allow some tuning ``parameters'' of the SP modules to be updated and optimized during training given a specific learning objective. 
%it is difficult to ; but we are also wasting the known partial information if we directly use DNN layers.

This paper develops a holistic approach of combining SP operators and DNN that overcomes the aforementioned challenges.  A hybrid neural network (hybrid-NN) is proposed, in which one or more properly-selected SP operators are inserted into the DNN architecture as embedded layers, and some key design parameters of each SP operator are treated as unknowns and updated during network training from data. To perform efficient training for the hybrid-NN, the (sub-)gradients of SP operators are utilized to compute the training error of each SP layer, which is then incorporated into the back-propagation method for iterative network training. The proposed training algorithm can
can simultaneously train the weights of the DNN and optimize the unknown tuning parameters of the SP operators from the labeled data. 
a%ctively involve the SP layers in the DNN training, such that both the weights of the DNN and the unknown tuning parameters of the SP operators can be  simultaneously optimized from the training data.  
The hybrid-NN offers a viable framework to take advantages of the strengths from both ordinary DNN and SP: the SP layers utilize the partially known models of the data to improve the sample efficiency in feature extraction, and the DNN architecture offers universality in learning the remaining unknown models and features from data. Simulation results on a radar ATR problem corroborate that the hybrid-NN offers enhanced capability in feature extraction, and can markedly reduce the amount of training data needed for DNN learning.  

%This paper is organized as follows. In Section \ref{sec:hnn}, the general idea of hybrid-NN framework is introduced. In order to test the feasibility of proposed framework, a demonstration application of hybrid-NN on radar automatic target recognition (ATR) problem is discussed in Section \ref{sec:atr}. The simulation result of hybrid-NN based radar ATR is given in Section \ref{sec:simu}, followed by the conclusion at the end.

\vspace{-.5em}
\section{Deep Neural Network}

\begin{figure}[t]
	\centering
	\includegraphics[width=.8\columnwidth]{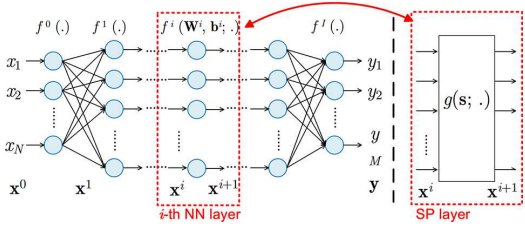}
	\vspace{-.5em}
	\caption{Diagram of (Hybrid) Neural Network}
	\label{fig:1}
	\vspace{-1.5em}
\end{figure}

%Layers Input Training
\vspace{-.5em}
Deep neural network \cite{lecun2015deep} is a highly structured framework. A typical DNN architecture is shown in the left part of Figure \ref{fig:1}, which is composed of many nonlinear processing stages, denoted as ``layers'', where each layer's output feeds to its immediate next layer as the input. For a DNN with $I$ layers, the relationship between input $\mathbf{x}^{i-1}$ and output $\mathbf{x}^{i}$ of a specific layer $i, 1\leq i\leq I,$ can be described as:
%\vspace{-.1em}
\begin{equation}
	\label{eq:1}
	\mathbf{x}^{i}=f^i(\mathbf{W}^i, \mathbf{b}^i; \mathbf{x}^{i-1})=f^i(\mathbf{W}^i\mathbf{x}^{i-1}+\mathbf{b}^i),
%	\vspace{-.5em}
\end{equation}
where $f^i(\cdot)$ is the activation function, $\mathbf{W}^i$ denotes the weights and $\mathbf{b}^i$ denotes the bias of the $i$-th layer. Usually $f^i(\cdot)$ is a non-linear function, e.g. sigmoid, tanh or ReLU \cite{lecun2015deep, glorot2011deep}.
%\vspace{-1em}

The entire network can be viewed as a cascade of function series $f^i(\cdot), i=1, \dots, I,$ as:
\vspace{-.1em}
\begin{equation}
	\label{eq:7}
	\begin{split}
		\mathbf{y}&=f^I(\mathbf{W}^I, \mathbf{b}^I; f^{I-1}(\mathbf{W}^{I-1}, \mathbf{b}^{I-1}; \dots f^1(\mathbf{W}^1, \mathbf{b}^1; \mathbf{x}^0))) \\
			&=f(\mathbf{W}, \mathbf{b}; \mathbf{x}^0),
	\end{split}
	\vspace{-.5em}
\end{equation}
where $\mathbf{x}^0$ is the input and $\mathbf{y}$ is the output of the entire network. The multi-layered DNN structure in (\ref{eq:7}) is attractive for its good universal function approximating ability \cite{csaji2001approximation}, which means that with a sufficient amount of training data, it is possible to use (\ref{eq:7}) to fit very complex problems via updating the parameters $(\mathbf{W}; \mathbf{b})$. 

The training stage usually adopts the back-propagation (BP) method \cite{lecun2015deep, rumelhart1986learning}. In each iteration of the BP method, the error of each layer is propagated backwards from the output layer to the input layer, and the update value of parameters in each layer is calculated simultaneously with the error propagation by calculating the gradient as:
\begin{equation}
	\Delta\mathbf{W}^i=-\lambda\left(\frac{\partial f^i}{\partial\mathbf{W}^i}\right)\boldsymbol{\delta}^i, \quad
	\boldsymbol{\delta}^{i-1}=\left(\frac{\partial f^i}{\partial\mathbf{x}^{i-1}}\right)\boldsymbol{\delta}^i,
\end{equation}
where $\Delta\mathbf{W}^i$ is the parameter update value of the $i$-th layer, $\boldsymbol{\delta}^i$ is the output error of the $i$-th layer (which is propagated from the $(i+1)$-th layer),   $\lambda$ is the learning rate and $\boldsymbol{\delta}^{i-1}$ is the error propagated back to the $(i-1)$-th layer.

%Usually $f^i(\cdot)$ is a non-linear function, e.g. sigmoid, tanh or ReLU \cite{}. There exist plenty of variants of NNs, for example, in convolutional neuron networks (CNNs) the weights are shared by some neurons, but the basic expression (\ref{eq:1}) still keeps invariant. Note that in many cases the data flows inside the network are matrices even tensors (e.g. multi-channel images, which forms 3-D tensors), but the basic idea remains the same.

\vspace{-.5em}
\section{Hybrid Neural Network}
\label{sec:hnn}
\vspace{-.5em}
%In this section, we propose the hybrid-NN architecture which combines the strengths of both SP and DNN. %conventional signal processing methods and neural network.

\vspace{-.5em}
\subsection{Hybrid Neural Network Design}
\vspace{-.5em}

%The above described NN framework has a good universal function approximating ability \cite{bibid}. 
In deep learning applications, the multi-layered framework in (\ref{eq:7}) provides excellent approximation ability in a wide range of problems. However, in some specific problems the layer model in (\ref{eq:1}) \emph{might not have the best approximation ability}.
%This often happens in the signal processing fields, where we usually deal with highly structured or man-made signals, which signals can be processed via mature signal processing methods with much more efficiency. 
This fact inspires us to design a specific \emph{signal processing (SP) layer} which is optimized for some special input data, and replace one (or some) layer(s) in the ordinary DNN to achieve a better feature extraction ability in some problems. The SP layer is designed following specific conventional SP operator based on some signal property that we want to utilize. Meanwhile, some parameters are not or cannot be pre-defined in the SP layer, but to be fine tuned in the training stage. % The cos ``universality'', which means the modified DNN only works for this type of data. 

Suppose that $\mathbf{x}^{i-1}$, the $(i-1)$-th layer's output of an ordinary DNN, can be effectively processed by a SP operator $p$ with parameters $\mathbf{s}$, but $\mathbf{s}$ is either unknown or needed to be fine tuned. Denoting the output of this SP method as $\tilde{\mathbf{x}}^{i}$, the relationship between $\mathbf{x}^{i-1}$ and $\tilde{\mathbf{x}}^{i}$ can be described as a mapping:  %$g: \mathrm{dim}(\mathbf{x}^{i-1})\mapsto\mathrm{dim}(\mathbf{x}^{i})$:
%A signal processing layer is given a
%\vspace{-.2em}
\begin{equation}
	\tilde{\mathbf{x}}^{i}=p(\mathbf{s}; \mathbf{x}^{i-1}).
%	\vspace{-.5em}
\end{equation}
Incorporating with an activation function $f^i(\cdot)$ (which can be chosen as any conventional DNN activation function), we can construct a network layer $g(\cdot)$ as:
\begin{equation}
	\label{eq:2}
	\mathbf{x}^{i}=g(\mathbf{s}; \mathbf{x}^{i-1})=f^i\left(p(\mathbf{s}; \mathbf{x}^{i-1})\right).
%	\vspace{-.5em}
\end{equation}
%where $g(\cdot)$ is the selected SP method and $\mathbf{s}$ is the parameter, as shown in figure \ref{fig:2}. 
We call (\ref{eq:2}) as the SP layer. We can insert this SP layer into an ordinary DNN by replacing one of its conventional layers. A DNN with its $i$-th layer replaced by (\ref{eq:2}) is expressed as:
%\vspace{-.2em}
\begin{equation}
	\label{eq:8}
	\begin{split}
		\mathbf{y}&=f^I(\mathbf{W}^I, \mathbf{b}^I; \dots g(\mathbf{s}; \dots f^1(\mathbf{W}^1, \mathbf{b}^1; \mathbf{x}^0))) \\
		&=f^\prime(\mathbf{W}, \mathbf{b}, \mathbf{s}; \mathbf{x}^0),
	\end{split}
%	\vspace{-1em}
\end{equation}
which is shown in the right part of Figure \ref{fig:1}. This modified DNN is named as hybrid neural network, or hybrid-NN in short.

The choice of mapping $p(\cdot)$ varies in applications in order to achieve the best feature extraction ability. For example, if $\mathbf{x}^{i-1}$ happens to be a BPSK signal (although we do not know its accurate parameters), the SP operator $p(\cdot)$ at the $i$-th layer can be chosen as a BPSK demodulator:
%\vspace{-.2em}
\begin{equation}
	p(\omega, \eta; \mathbf{x}^{i-1})=\mathrm{conv}(F(\eta), \mathbf{r}(\omega)^\mathrm{T}\mathbf{x}^{i-1}),
%	\vspace{-.5em}
\end{equation}
where $\mathrm{conv}(\cdot)$ is the convolution operator, $F(\eta)$ is a low-pass filter parameterized by $\eta$ and $\mathbf{r}(\omega)$ is the reference signal with carrier frequency $\omega$. Key to this SP layer is that we adopt the convolution operator to make use of the model structure of an optimal BPSK demodulator, and at the same time allow the key parameters $\mathbf{s}:=(\eta, \omega)$ to be unknown \emph{a priori}, such that they can be learned during training. 

%In learning problems we usually do not know the exact values of $\mathbf{s}$, which is needed to be learned from the data. But we do know the $\mathbf{x}^i$ exploits some known features and can be effectively processed via a specific SP method, which can be used to design a proper $g(\cdot)$. %For example, in a detection problem, you may expect a matched filter as $g(\cdot)$, while leave the parameters of the filter to be learned during the training stage. 

%The entire hybrid-NN framework architecture is shown in figure \ref{fig:2}.
%\begin{figure}[h]
%	\centering
%	\includegraphics[width=.6\columnwidth]{2.eps}
%	\caption{Diagram of SP Layer in Hybrid Neural Network}
%	\label{fig:2}
%\end{figure}

\vspace{-.5em}
\subsection{Hybrid Neural Network Training Algorithm}
\label{sec:train}
\vspace{-.5em}

As a novel DNN architecture, we develop the training algorithm of hybrid-NN in this subsection.
%There are some issues needed to be addressed during the implementation of a hybrid-NN.
%
%\vspace{-.5em}
%\subsubsection{Training Algorithm}
%\vspace{-.5em}
Since the overall architecture of hybrid-NN is still a layer-wise structure, we can adopt the conventional BP method \cite{lecun2015deep, rumelhart1986learning} to train it. The key is to deal with the SP layer(s): how to update the parameters $\mathbf{s}$ in SP layer, and how to back-propagate the error to its neighboring layer.

Assume that the partial derivatives of the SP operator in (\ref{eq:2}) exist. Given the $i$-th layer is SP and the output error of this layer is:
%\vspace{-.2em}
\begin{equation}
	\label{eq:a1}
	\boldsymbol{\delta}^i=\boldsymbol{\delta}^{i+1} f^{i\prime}\left( p(\mathbf{s}; \mathbf{x}^{i-1})\right),
%	\vspace{-.5em}
\end{equation}  
where $\boldsymbol{\delta}^{i+1}$ is the error from $(i+1)$-th layer and $f^{i\prime}$ is the derivative of the activation function. Then, the update value of $\mathbf{s}$ and the output error back-propagated to previous layer can be calculated by gradient descent as:
%\vspace{-.2em}
\begin{equation}
	\label{eq:a2}
	\Delta\mathbf{s}=-\lambda\left(\frac{\partial p}{\partial\mathbf{s}}\right)\boldsymbol{\delta}^i, \quad
%\end{equation}
%\begin{equation}
	\boldsymbol{\delta}^{i-1}=\left(\frac{\partial p}{\partial\mathbf{x}^{i-1}}\right)\boldsymbol{\delta}^i,
%	\vspace{-.5em}
\end{equation}
where $\lambda$ is the learning rate.
%$\boldsymbol{\delta}^{i-1}$ is previous layers as ordinary BP does.

\noindent\emph{Remark 1: Complex values.}
Complex values are inevitable in hybrid-NN because complex-valued signals are common in signal processing problems.
%, hence the complex-valued neural networks (CVNNs)  are inevitable if we want to combine SP and DNN. 
%For complex-valued neural networks (CVNNs) \cite{hirose2012complex}, conventional gradient descent based BP training methods are no longer available. Various CVNN frameworks and corresponding training algorithms are developed \cite{georgiou1992complex, gangal2009inversion, nitta2009complex}, which can be directly adopted to complex-valued hybrid-NN. 
One straightforward training approach is to calculate the gradient in (\ref{eq:a1}) and (\ref{eq:a2}) using Wirtinger calculus \cite{amin2011wirtinger}. 
%Suppose the (complex-valued) output error of the SP layer is $\boldsymbol{\delta}^k$, the update value of the weights is:
%%\vspace{-.2em}
%\begin{equation}
%	\Delta\mathbf{W}^k=-\lambda\left(\mathbf{J}_{\mathbf{W}^k}^\mathrm{H}\boldsymbol{\delta}^k+(\mathbf{J}_{\mathbf{W}^k}^\mathrm{\ast H}\boldsymbol{\delta}^k)^\ast\right)\mathbf{x}^{k-1},
%%	\vspace{-.5em}
%\end{equation}
%where $\mathbf{J}_{\mathbf{W}^k}$ and $\mathbf{J}_{\mathbf{W}^k}^\ast$ are two Jacobian matrices defined via Wirtinger calculus:
%%\vspace{-.2em}
%\begin{equation}
%	\mathbf{J}_{\mathbf{W}^k}=\frac{\partial f^k}{\partial\mathbf{W}^k}, \quad\mathbf{J}_{\mathbf{W}^k}^\ast=\frac{\partial f^k}{\partial(\mathbf{W}^{k})^\ast}.
%%	\vspace{-.5em}
%\end{equation}
%The rest of training steps remain the same as BP in real-valued DNN.

\noindent\emph{Remark 2: Subgradient.}
In case the SP operator in (\ref{eq:2}) is not differentiable, we can still use (\ref{eq:a1}) and (\ref{eq:a2}) by replacing the derivatives by subderivatives.

\vspace{-.5em}
\noindent\emph{Remark 3: Size of the training dataset.}
Compared with the ordinary DNN, a hybrid-NN only works for specific problems that the SP method $p(\cdot)$ is suitable for. As the reward, SP layer is expected to have better feature extraction ability, which in turn reduces the number of training iterations. Furthermore, usually the size of $\mathbf{s}$ is much smaller than the size of $(\mathbf{W}^i; \mathbf{b}^i)$ in an ordinary DNN layer, which means the number of unknown parameters can be reduced and we can use less data to train it. %This is extremely attractive in SP problems e.g. channel estimation where the available training data amount is usually limited.

\vspace{-.5em}
\subsection{Discussions}
\vspace{-.5em}
\subsubsection{SP Layer Placement}
In general, the location index $i$ of the SP layer can be any value between $1$ and $I$ which is the number layers in the neural network. But the SP layer usually works effectively for structured or modeled input data. Currently, the interpretation of data inside hidden layers of an ordinary DNN (which means $2\leq i\leq I-1$) is still a huge challenge, making us difficult to find a suitable SP operator. %Without a good understanding of the signals in the hidden layers, it is very difficult for us to find a suitable SP tool. 
The most well-understood data of DNN are its inputs and outputs, i.e. $\mathbf{x}^0$ and $\mathbf{y}$, which means that at current stage we prefer to put the SP layer at the beginning or the end of a hybrid-NN.

\vspace{-.5em}
\subsubsection{Difference From Pre-/Post-processing}
\vspace{-.5em}
Compared with existing work of combing SP and DNN such as \cite{PCAnet, Optnet}, the SP operator in in the hybrid-NN is no longer a \emph{hyper-parameter} that is determined before training. In fact, it can be any (sub-)differentiable SP operator with unknown $\mathbf{s}$.
We leave $\mathbf{s}$ as a parameter of hybrid-NN to learn it from data, which provides us with enhanced design flexibility compared with existing work. 

%the SP operator in hybrid-NN is no longer a \emph{hyper-parameter} which is determined before training. SP layer of hybrid-NN can be any (sub-)differentiable SP operator, and we leave $\mathbf{s}$ as a \emph{parameter} of hybrid-NN to learn it from data. This provides us much more flexibility compared with existing work. 

\section{Application of Hybrid-NN in Radar Automatic Target Recognition}
\label{sec:atr}

\begin{figure}[t]
	\centering
	\subfigure[Conventional ATR Using DNN]{
		\label{fig:4a}
		\includegraphics[width=.7\columnwidth]{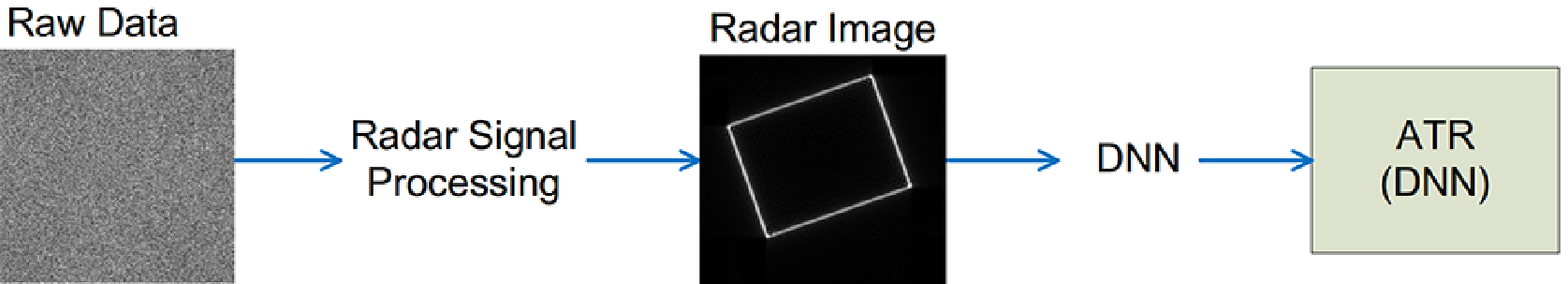}
	}\\
	\subfigure[Proposed ATR Using Hybrid-NN]{
		\label{fig:4b}
		\includegraphics[width=.7\columnwidth]{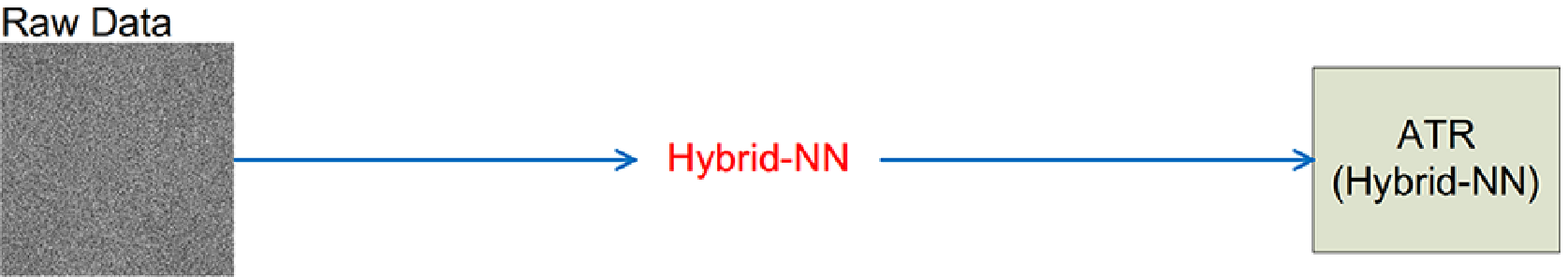}
	}
	\caption{Radar ATR Frameworks}
	\label{fig:4}
	\vspace{-.1em}
\end{figure}

In order to demonstrate the feasibility of hybrid-NN, we show an example in radar automatic target recognition (ATR).
ATR refers to identifying and classifying the targets automatically from the received data, in contrast to the traditional human-aided target recognition. State-of-the-art DNN-based radar ATR techniques use a two-stage approach, which firstly generates the radar image from the raw radar signal via signal processing algorithms 
%such as range-Doppler (RD), chirp scaling (CS) or $\omega-k$ algorithm 
\cite{cumming2005digital}, and then perform automatic classification using DNN methods
%. Traditionally, automatic classification is difficult using conventional SP methods due to the complexity of target scenes, but DNN based methods have been successfully applied 
\cite{morgan2015deep, chen2016target, wilmanski2016complex}, as described in Figure \ref{fig:4a}. Unfortunately, such SP-based pre-processing is not always applicable in practice, because some radar parameters are either unavailable or unreliable which makes the traditional radar signal processing impossible.

In this paper, we suggest a novel ATR framework as shown in Figure \ref{fig:4b}, which directly performs automatic classification from the raw radar signals using hybrid-NN, bypassing the radar signal processing stage. The SP and DNN are coherently combined, which is particularly attractive for real-time ATR, as well as in situations where some radar parameters need to be learned or tuned. % from the training data. 

%This is particularly attractive for real-time ATR, %, because usually doing real-time radar signal processing on the vehicle is expensive due the resource limitation, but exploiting a well-trained DNN is cheaper. 
%as well as the case that some radar parameters are unavailable or unreliable which makes the traditional radar signal processing impossible, but we do have labeled training data to learn them out.  

\vspace{-.5em}
\subsection{Radar Model}
\vspace{-.5em}
Consider a simple radar signal model, where a radar transmits chirp waveforms to a point target which is uniformly moving in a straight line with backscattering coefficient $\sigma$. The received baseband echo is given in \cite{cumming2005digital}:
\begin{equation}
	\label{eq:3}
		s(\tau, t)\approx \sigma \exp(j\pi K_r (\tau-\tau_0)^2) \exp(-j\pi K_a (t-t_0)^2),
\end{equation}
%\begin{equation}
%	p(\tau)=\exp(j\pi K_r \tau^2),
%\end{equation}
%where $K_r$ is the range frequency modulation rate and $\tau$ is the range time. For a linear uniformly moving target with backscattering coefficient $\sigma$ and moving speed $v$, the baseband received echo can be approximated as
%\begin{equation}
%	\label{eq:3}
%	\begin{split}
%		s(\tau, t)&=\sigma \exp\left(j\pi K_r \left(\tau-\frac{2R(t)}{c}\right)^2\right) \exp\left(-\frac{4j\pi f_c R(t)}{c}\right)\\
%			&\approx \sigma \exp(j\pi K_r (\tau-\tau_0)^2) \exp(-j\pi K_a (t-t_0)^2),
%	\end{split}
%\end{equation}
%where $t$ is the azimuth time, $R(t)=\sqrt{R_0+v^2t^2}$ is the instantaneous range distance between radar and target, $R_0$ is the nearest range, $f_c$ is the carrier frequency, $\tau_0=2R_0/c$ is the range time delay, $K_a=\frac{2f_c v^2}{cR_0}$ is the azimuth frequency modulation rate and $t_0$ is the azimuth time that the target passes through the nearest point.
%For a target with multiple inputs, the echo is summation of each point.
where $K_r$ is the range frequency modulation rate, $\tau$ is the range time, $K_a$ is the azimuth frequency modulation rate, $t$ is the azimuth time, $\tau_0$ is the range time delay and $t_0$ is the azimuth time that the target passes through the nearest point of the target moving trajectory with respect to the radar. %In real applications, $K_r$ and $K_a$ are unknown, hence conventional ATR methods do not work. 

The raw radar data is stored as a 2-D data matrix $\mathbf{S}$ whose entries are digital samples of $s(\tau, t)$ along $\tau$ and $t$, with range sampling rate $F_\mathrm{sr}$ and pulse repetition frequency (azimuth sampling rate) $\mathrm{PRF}$. These sampling rates are selected by user. 

\vspace{-.5em}
\subsection{Architecture of Hybrid-NN for Radar ATR}
\vspace{-.5em}
Radar ATR is basically a classification problem.
Given labeled training data, it is straightforward to train a universal convolutional neural network (CNN) for target classification, in the absence of any knowledge of the signal model in (\ref{eq:3}). Alternatively, our hybrid-NN approach is make use of (\ref{eq:3}) for improved efficiency in training and learning. Our key step is to design a suitable radar signal processing layer, and insert this layer into a proper location of an ordinary CNN, proposing a hybrid-NN for radar ATR from raw data.

Conventional SP algorithms for radar are based on matched filtering (MF). The corresponding matched filter for (\ref{eq:3}) is:
\begin{equation}
	\label{eq:4}
	m(\tau, t)=\exp\left(-j\pi \hat{K}_r \tau^2 \right)\exp\left(j\pi \hat{K}_a t^2\right).
\end{equation}
Ideally, the MF parameters $(\hat{K_a}，\hat{K_r})$ are determined by $K_a$ and $K_r$. But in real applications, $K_a$ and $K_r$ are either unknown (e.g. passive radar) or inaccurate due to platform and system errors, hence $(\hat{K_a}，\hat{K_r})$ are needed to be trained.

Accordingly, we design a \emph{MF layer} as:
\begin{equation}
	\label{eq:5}
	\begin{split}
		g(\hat{K_a}, \hat{K_r}; \mathbf{S})&=f\left(\mathrm{conv}\big\{\mathbf{M}(\hat{K_a}, \hat{K_r}), \mathbf{S}\big\}\right)\\
		&=\left|\mathrm{conv}\big\{\mathbf{M}(\hat{K_a}, \hat{K_r}), \mathbf{S}\big\}\right|,
	\end{split}
\end{equation}
where the activation function $f(\cdot)$ is chosen as the absolute value function $|\cdot|$, $\mathbf{S}$ is the radar raw data and $\mathbf{M}(\hat{K_a}, \hat{K_r})$ is the matched filter defined as (\ref{eq:4}). The $(m, n)$-th element of $\mathbf{M}$ is:
\begin{equation}
	\label{eq:6}
	\left[\mathbf{M}(\hat{K_a}, \hat{K_r})\right]_{mn}=e^{-j\pi \hat{K_r} m^{\prime 2} }e^{j\pi \hat{K}_a n^{\prime 2}},
\end{equation}
where $m^\prime=\frac{m}{F_\mathrm{sr}}$ and $n^\prime=\frac{n}{\mathrm{PRF}}$. 

\begin{figure*}[th!]
	\centering
	\includegraphics[width=.73\textwidth]{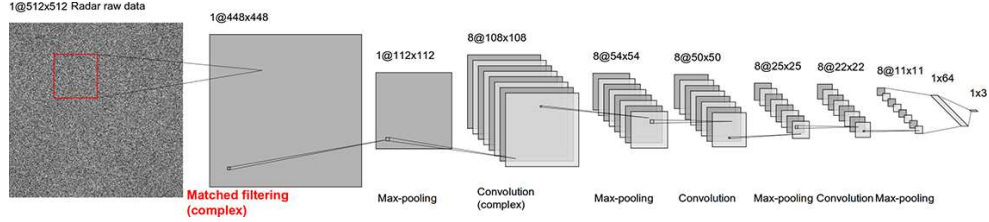}
	\caption{Architecture of Hybrid-NN for Radar ATR}
	\label{fig:9}
\end{figure*}
The MF layer is used as the first layer of the network and followed by an ordinary CNN. The concept is intuitive: MF layer is capable of utilizing the known properties of the radar data in terms of radar waveform structure, and then the extracted information by MF is fed into an ordinary CNN to do classification. During the training stage, the SP-layer parameters $\mathbf{s}:=(\hat{K}_a, \hat{K}_r)$ will be updated automatically through the BP algorithm presented in Section 3.2. 
The configuration of the hybrid-NN is shown in Figure \ref{fig:9}. The network is composed of one matched filtering layer, three convolutional layers ($5\times 5\times 8, 5\times 5\times 8$ and $4\times 4\times 8$) and one fully connected layer ($64$). The size of matched filter is $64\times 64$. For an ordinary convolutional layer, a $64\times 64$ kernel has $4,096$ parameters to learn, but for a matched filtering layer we only have two parameters $\hat{K_a}$ and $\hat{K_r}$ to determine. This provides us a great reduction on the amount of training data. %The convolutional layers are with sizes of $5\times 5\times 8, 5\times 5\times 8$ and $4\times 4\times 8$ respectively, followed by a fully connected layer of size $64$.
The network is trained using the algorithm given in Subsection \ref{sec:train}.

\vspace{-.5em}
\section{Simulations}
\label{sec:simu}
\vspace{-.5em}
\subsection{Data Generation}
\begin{figure}[h!]
	\centering
	\subfigure[Raw Data]{
		\includegraphics[width=.2\columnwidth]{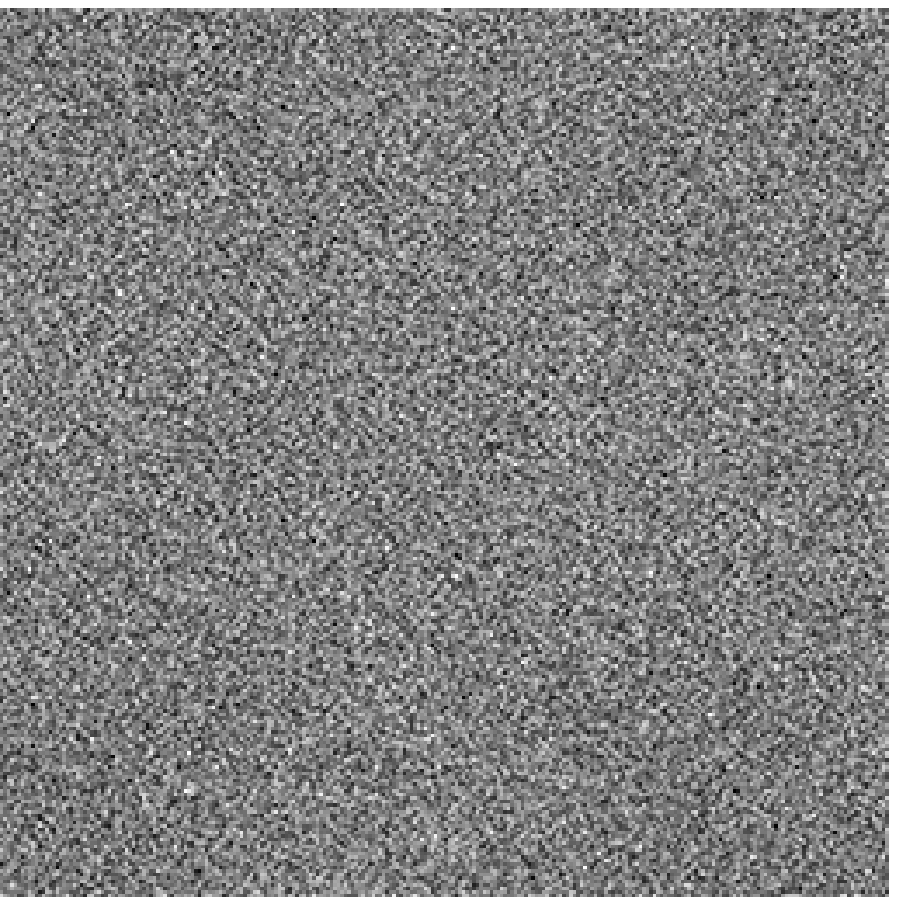}
	}
	\subfigure[Circle]{
		\includegraphics[width=.2\columnwidth]{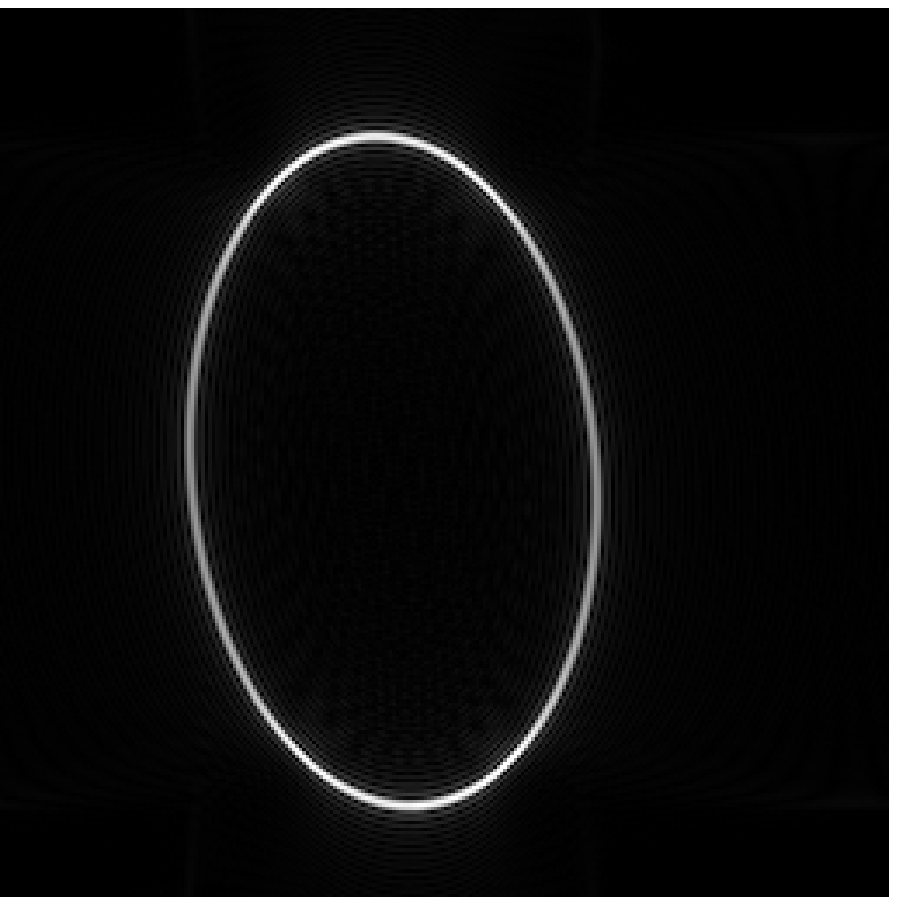}
	}
	\subfigure[Square]{
		\includegraphics[width=.2\columnwidth]{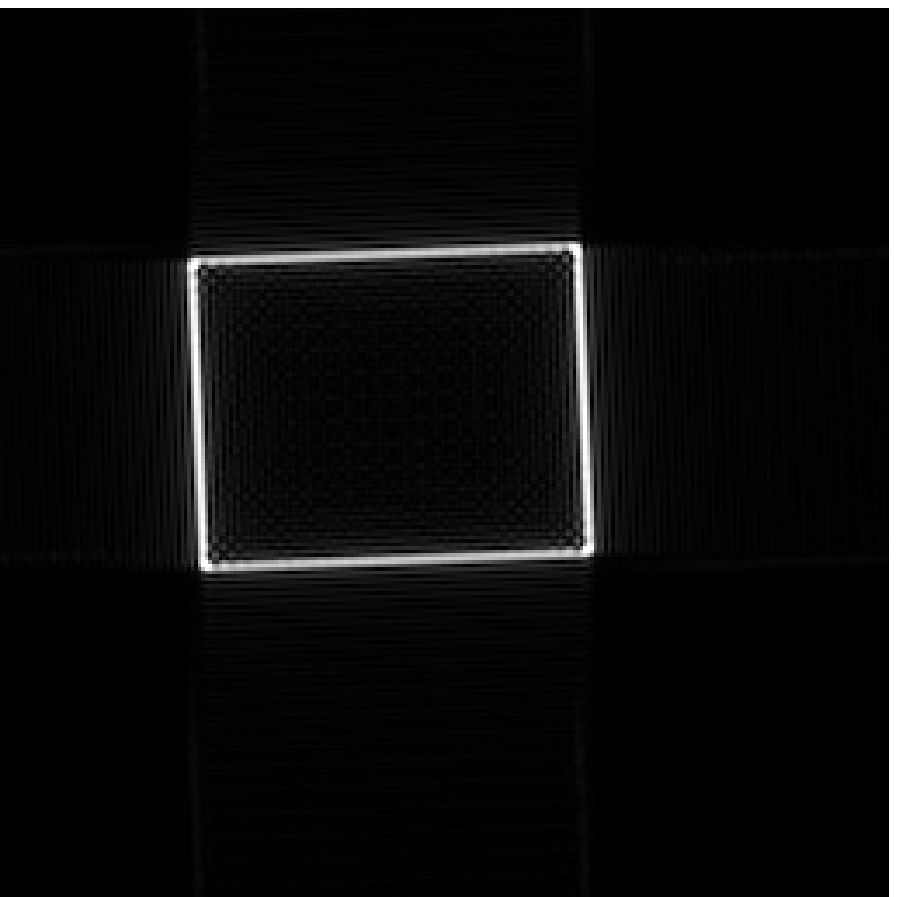}
	}
	\subfigure[Triangle]{
		\includegraphics[width=.2\columnwidth]{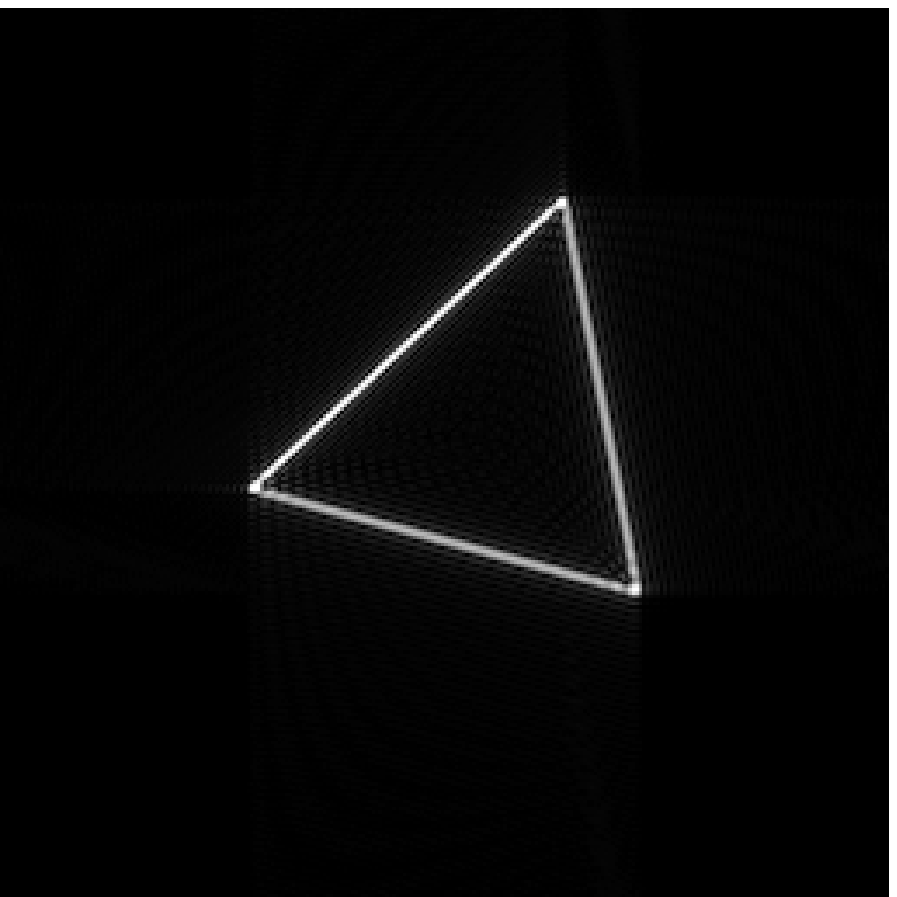}
	}
	\caption{Samples Data for Radar ATR}
	\label{fig:3}
\end{figure}

\vspace{-.5em}
Simulation data is generated to train and validate the performance of the proposed hybrid-NN for radar ATR. The training set includes three types of targets: circles, squares and triangles which are generated with random magnitudes, random deformations and random noises, as shown in Figure \ref{fig:3}. The radar parameters are given in Table \ref{table:1}. 
Also note that radar raw data size are usually very large, here set as $512\times 512$. Such a large input size incurs tremendous computational load to the training stage of ordinary DNN.
\vspace{-.5em}
\vspace{-3em}
\begin{table}[h]
	\centering
	\caption{Simulation Parameters}
	\label{table:1}
	\small
	\begin{tabular}{ll}
		\hline
		Carrier Frequency & 5 GHz \\
		Range Sampling Rate & 600 MHz \\
		Pulse Duration & $10 \mathrm{\mu s}$ \\
		Range Bandwidth & 500 MHz \\
		Range Distance & 5,000 m \\
		Target Speed & 100 m/s \\
		PRF & 1,000 Hz \\
		\hline
	\end{tabular}
\end{table}

%\subsection{Hybrid-CNN Implementation}
%
%% Radar data amount is usually very huge. 
%
%The configuration of hybrid-CNN used in the simulation is shown in Figure \ref{fig:9}. The detailed configurations of each layer is listed in Table \ref{table:2}.
%
%\begin{figure*}[htbp]
%	\centering
%	\includegraphics[width=.9\textwidth]{10.eps}
%	\caption{Architecture of Hybrid-CNN for Radar ATR}
%	\label{fig:9}
%\end{figure*}
%\begin{table}[h]
%	\caption{Layers in Proposed Hybrid-CNN}
%	\label{table:2}
%	\small
%	\begin{tabular}{ll}
%		\hline
%		Layer & Dimensionality \\
%		\hline
%		Input layer & $512\times 512\times 1$, complex \\
%		Matched filtering & $64\times 64\times 1$, complex \\
%		Max-pooling (on amplitude) & $4\times 4$ \\
%		Convolution / Abs ($|\cdot|$) & $5\times 5\times 8$, complex \\   
%		Max-pooling & $2\times 2$ \\
%		Convolution / ReLU & $5\times 5\times 8$ \\
%		Max-pooling & $2\times 2$ \\
%		Convolution / ReLU & $4\times 4\times 8$ \\
%		Max-pooling & $2\times 2$ \\
%		Fully connected / ReLU & 64 \\
%		Fully connected / Softmax & 3 \\
%		\hline
%	\end{tabular}
%\end{table}
\vspace{-2em}

\subsection{Numerical Results}
\vspace{-.5em}
As a benchmark, an ordinary complex-valued DNN (which is in fact a CNN here) is adopted with similar architecture as the described hybrid-NN but only switches its first layer to a convolutional layer. %The configuration of the rest parts of the network is the same as hybrid-CNN described in previous subsection.

\begin{figure}[hb]
	\centering
%	\subfigure[Training Cost Function]{
%		\includegraphics[width=.7\columnwidth]{fig/7.eps}
%	}\\
%	\subfigure[Training Accuracy]{
		\includegraphics[width=.7\columnwidth]{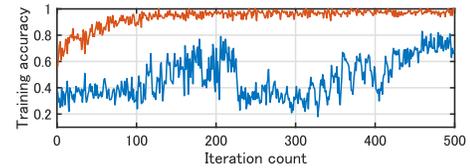}
%	}
	\caption{Training Accuracy of Hybrid-NN and ordinary DNN with 5,000 images.}
	\label{fig:6}
\end{figure}

First, the training performance of hybrid-NN compared with conventional DNN are shown in Figure \ref{fig:6}. A training set with 5,000 images is used, with each $\mathrm{mini\_batch}$ of 50 images and trained for 5 epochs\footnote{Number of counts that the entire training dataset is used once.}. This is a very small training set, especially for the large input size. The proposed hybrid-NN shows great advantage on the training performance. It can be seen that during the 4th epoch, the hybrid-NN has already converged to a good optimum whereas the ordinary DNN cannot converge yet. Finally, hybrid-NN ends with 98\% training accuracy, compared with 64\% of the ordinary DNN. In order to determine the data requirement of the ordinary DNN, the dataset size is further increased to 25,000 images, and the network finally converged with 82\% accuracy as shown in Figure \ref{fig:9}. This simulation verifies the benefits of hybrid-NN in the training stage, including fast convergence and small training dataset size.

\begin{figure}[hb]
	\centering
	\includegraphics[width=.7\columnwidth]{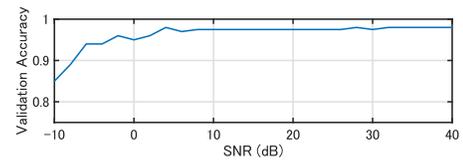}
	\caption{Validation Accuracy with Different SNR}
	\label{fig:8}
\end{figure}

Second, the accuracy of trained hybrid-NN on the validation data with different noise levels is also tested in Figure \ref{fig:8}. As the SNR is changed from -10dB to 40dB, the validation accuracy starts with 85\% at -10dB and rapidly grows to 96\% after 0dB. This result shows the robustness of proposed hybrid-NN and very high validation accuracy on the test data in the validation stage.

%\subsection{Discussions}
\vspace{-1em}
\section{Conclusion}
\label{sec:con}
\vspace{-.5em}
We introduce a novel hybrid-NN framework, which inject the DNN with a SP layer that is specifically designed for particular signal models. A network training algorithm is presented to simultaneously update both the network weights and the design parameters of the SP layer during training. The proposed hybrid-NN framework is tested on a radar ATR application. Compared with ordinary DNN, the proposed hybrid-NN dramatically reduces the required amount of training data and improves the training efficiency with high validation accuracy. % Future works on extending this framework to more applications and performance analysis are planned.
\vspace{-1em}
\begin{figure}[hb]
	\centering
	%	\subfigure[Training Cost Function]{
	%		\includegraphics[width=.7\columnwidth]{fig/7.eps}
	%	}\\
	%	\subfigure[Training Accuracy]{
	\includegraphics[width=.7\columnwidth]{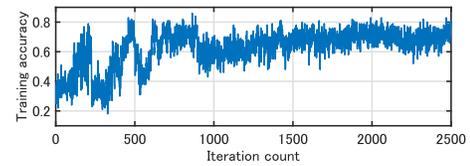}
	%	}
	\caption{Training Accuracy of ordinary DNN with 25,000 images}
	\label{fig:9}
\end{figure}
\vspace{-1em}
\vspace{-1em}
%\vspace{-.5em}
% Below is an example of how to insert images. Delete the ``\vspace'' line,
% uncomment the preceding line ``\centerline...'' and replace ``imageX.ps''
% with a suitable PostScript file name.
% -------------------------------------------------------------------------

%\vfill
%\pagebreak

%\vfill\pagebreak
%\section{REFERENCES}
%\label{sec:refs}
%\newpage

\newpage

\bibliographystyle{IEEEbib}
\bibliography{ref}

\end{document}